\documentclass[doublecol]{epl2} 

\usepackage[normalem]{ulem}

\title{Non-Ergodicity in open quantum systems \\ through quantum feedback}
\shorttitle{Non-Ergodicity in open quantum systems through quantum feedback} 

\author{Lewis A.~Clark,\inst{1,4} Fiona Torzewska,\inst{2,4} Ben Maybee\inst{3,4} and Almut Beige\inst{4}}
\shortauthor{L. A. Clark \etal}

\institute{                    
  \inst{1} Joint Quantum Centre (JQC) Durham-Newcastle, School of Mathematics, Statistics and Physics, Newcastle University, Newcastle-Upon-Tyne, NE1 7RU, United Kingdom\\
  \inst{2} The School of Mathematics, University of Leeds, Leeds LS2 9JT, United Kingdom\\
  \inst{3} Higgs Centre for Theoretical Physics, School of Physics and Astronomy, The University of Edinburgh, Edinburgh EH9 3JZ, UK, United Kingdom\\
  \inst{4} The School of Physics and Astronomy, University of Leeds, Leeds LS2 9JT, United Kingdom\\
}
\pacs{42.50.-p}{Quantum optics}
\pacs{42.50.Ar}{Photon statistics and coherence theory}
\pacs{42.50.Lc}{Quantum fluctuations, quantum noise, and quantum jumps}

\abstract{It is well known that quantum feedback can alter the dynamics of open quantum systems dramatically. In this paper, we show that non-Ergodicity may be induced through quantum feedback and resultantly create system dynamics that have lasting dependence on initial conditions. To demonstrate this, we consider an optical cavity inside an instantaneous quantum feedback loop, which can be implemented relatively easily in the laboratory. Non-Ergodic quantum systems are of interest for applications in quantum information processing, quantum metrology and quantum sensing and could potentially aid the design of thermal machines whose efficiency is not limited by the laws of classical thermodynamics.}

\begin{document}

\maketitle

\section{Introduction}
Looking at the mechanical motion of individual particles, it is hard to see why ensembles of such particles should ever thermalise. Every particle moves in a well-defined manner through its available phase space while interacting from time to time, for example via elastic collisions, with other members of the ensemble. All equations of motion obey time-reversal symmetries and the information about initial phase space locations is never lost. Although individual particles fail to thermalise, macroscopic observables of physical systems usually tend towards constant values, as if the ensemble reached a well-defined thermal state. The key to understanding this seeming contradiction is {\em Ergodicity}. Ergodic systems tend to reach a stationary state, where time averages are well defined.  In such a case, these become equivalent to ensemble averages, which lie at the heart of statistical physics \cite{Boltzmann,Boltzmann2}.   

The emergence of Ergodicity and non-Ergodicity in quantum physics is still the subject of current debate. By 1929, von Neumann had proved the Ergodicity of finite-sized closed quantum mechanical systems, which evolve according to a Schr\"odinger equation, thereby verifying the applicability of statistical-mechanical methods to quantum physics \cite{Boltzmann3}. Moreover, it has been shown that open quantum systems that eventually lose any information about their initial state and whose dynamics result in a unique stationary state are Ergodic \cite{Kampen,Cresser,Mandel}. For isolated many-body quantum systems without an external bath, the eigenstate thermalisation hypothesis \cite{Mark,many,Mark2} has been introduced to explain when and why they can be described by equilibrium statistical mechanics.

Although the dynamics of open quantum systems are in general Ergodic, many classical stochastic processes are not, particularly in condensed matter systems \cite{Broken}.  Hence if classical physics emerges from microscopic quantum models, there have to be mechanisms which induce non-Ergodicity.  Because of this, Ergodicity breaking is an active area of research in a variety of quantum disciplines \cite{Ines,Papic,Papic2,Matin} In this paper, we discuss an example of such a mechanism and show that it can be used to induce non-Ergodicity in quantum optical systems with spontaneous photon emission.

For open quantum systems, which evolve with a Lindbladian master equation, non-Ergodicity seems to require the existence of multiple stationary states which typically occur only in carefully designed circumstances \cite{Kampen,Cresser,Mandel} e.g.~in systems with mutiple symmetries and degeneracies in their dynamics.  Moreover, non-Ergodicity of open quantum systems can be caused by interaction with non-harmonic thermal environments \cite{Ines}.  In addition, some isolated (i.e.~closed) many-body quantum systems which evolve according to the Schr\"odinger equation and can be studied numerically have been shown to exhibit non-Ergodic dynamics after sudden quenches \cite{many,Mark2}. Very recently, a new form of weak Ergodicity breaking with a range of unusual properties which violate the eigenstate thermalisation hypothesis has been shown to occur in constrained quantum systems and has been interpreted as a manifestation of quantum many-body scars \cite{Papic,Papic2}.

In this paper, we identify another mechanism with the ability to induce non-Ergodic dynamics in open quantum systems without the need for multiple stationary states.  More concretely, we show that open quantum systems can become non-Ergodic in the presence of sufficiently strong quantum feedback \textemdash even when the system possesses only a single well-defined stationary state. This constitutes an interesting observation, since non-Ergodicity and non-linear dynamics provide useful tools for processing quantum information \cite{HQMM,HQMM2}, quantum metrology and sensing \cite{PRA,PRA2}. Moreover, it could aid the design of thermal machines which are not limited by the laws of classical thermodynamics \cite{Jaynes,Jaynes_book}. 

\begin{figure}[t]
\centering
\includegraphics[width=0.49 \textwidth]{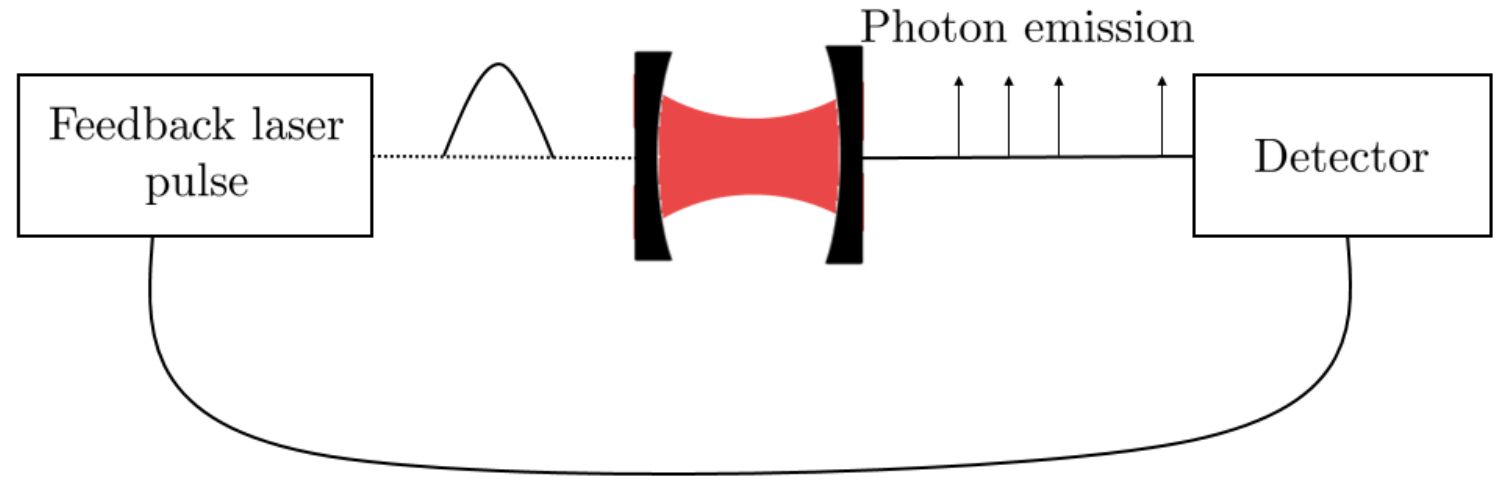}
\caption{[Colour online] Schematic view of an optical cavity inside an instantaneous quantum feedback loop.  A cavity field is prepared in a coherent state and allowed to freely decay.  Upon detection of a photon, a feedback pulse is applied to the cavity on a very short timescale that can be considered instantaneous.  This pulse has the effect of displacing the cavity field and as such alters its photon emission statistics.} \label{Setup2}
\end{figure}

As we shall see below, the individual trajectories of an open quantum system can exhibit different types of dynamics in the presence of quantum feedback, even when initialised in the same state. For example, some quantum trajectories might reach a fixed point of their dynamics, while others might evolve further and further away from it. If the feedback is strong enough, the stationary state of an open quantum system can become repulsive. Instead of losing any information about the initial state, there can be a persistent dependence of ensemble averages in expectation values on initial conditions. This differs from the standard use of quantum feedback as a tool to stabilise dynamics \cite{Wiseman-Milburn,Schirmer,Stable1,Stable2}.  To illustrate this, we study the dynamics of an optical cavity with spontaneous photon emission inside an instantaneous quantum feedback loop, as shown in Fig.~\ref{Setup2}.  

\section{Ergodicity} Let us first have a closer look at how to define Ergodicity in open quantum systems.  There are a number of ways to define Ergodicity in the literature, with equal validity \cite{Zelditch}.  In a number of quantum systems, Ergodicity is considered to be the ability to occupy the entire phase space.  If the dynamics are restricted to only a portion of this phase space, this would be a non-Ergodic process.   In the following, we instead take a definition which is based on classical statistical mechanics, motivated by our interest in the emergence of classical physics from quantum processes.  Suppose a large number of individual systems generate time-dependent stochastic signals. In the following, we call the dynamics of these systems Ergodic when any single, sufficiently long sample of the process has the same statistical properties as the entire process.  Let us consider $N$ identically prepared systems with stochastic dynamics and denote the ensemble average of the individual expectation values $E_n(A,T)$ of an observable $A$ after a long time $T$ by $E(A)$, which implies that 
\begin{eqnarray} \label{ergodicity2}
E(A) &=&  \lim_{T \to \infty} \lim_{N \to \infty} {1 \over N} \sum_{n=1}^N E_n(A,T) \, .
\end{eqnarray}
Using this notation, the system dynamics are Ergodic when
\begin{eqnarray} \label{ergodicity}
E(A) &=& \lim_{T \to \infty} {1 \over T} \int_0^T {\rm d}t \, E_n(A,t) 
\end{eqnarray}
also, for all observables $A$ and for all systems $n$. In other words, system dynamics are Ergodic when ensemble averages and time averages are equivalent for all observables and for all possible process realisations \cite{Kampen}.

Within this paper, we will have a closer look at the possible trajectories of open quantum systems with Markovian dynamics. The density matrix $\rho$ of such systems, which represents the quantum state of an ensemble, can always be described by a Lindbladian master equation \cite{Lindblad,Wiseman-Milburn}.  For a system with only a single decay channel, this equation is of the form
\begin{eqnarray} \label{Superoperator}
\dot{\rho} & = & - \frac{\rm i}{\hbar} \left[H, \rho\right] + \frac{1}{2} \Gamma \left(2 L \rho L^\dagger - \left[L^\dagger L , \rho \right]_+ \right) ,~~
\end{eqnarray}
where $H$ is the system Hamiltonian, $\Gamma$ denotes a spontaneous decay rate and $L$ is a so-called Lindblad operator. The operator $L$ might simply be an energy annihilation operator in a system that freely decays. Consequently, the time derivative of the expectation value $\langle A \rangle = {\rm Tr}(A \rho)$ of an observable $A$ equals
\begin{eqnarray} \label{Superoperator2}
\langle {\dot A} \rangle & = & - \frac{\rm i}{\hbar} \left \langle \left[A,H \right] \right \rangle 
+ \frac{1}{2} \Gamma \left \langle L^\dagger \left[ A, L \right] + \left[ L^\dagger, A \right] L \right \rangle . ~~~
\end{eqnarray}
Usually the last term in this equation simplifies to the expectation value of an operator which equals $A$ or is at worst a relatively simple function of $A$.

When adding quantum feedback, the system dynamics still obey a Lindblad master equation but $L$ is now the product of a unitary operator and the unperturbed Lindblad operator of the system.  When this applies, the latter terms in the above equation may not simplify, even if they do so in the absence of quantum feedback. As a result, the dynamics of $\langle A \rangle$ may be governed by an infinitely large set of linear differential equations, capable of effectively generating non-linear dynamics. The accompanying non-Ergodicity that we discuss in this paper is thus qualitatively different from the non-Ergodicity in other open quantum systems \cite{Ines} and from that in localised many-body quantum systems with Hamiltonian dynamics \cite{many,Mark2,Papic,Papic2}. Moreover, it has been shown already that the mechanism which we discuss here can lead to applications in quantum technology, especially in quantum metrology \cite{PRA,PRA2}.

\section{An optical cavity with quantum feedback} To determine whether a quantum system is Ergodic or not, we need to compare the dynamics of its ensemble averages with the dynamics of its individual quantum trajectories. As an example, we now have a closer look at the experimental setup which is shown in Fig.~\ref{Setup2} and consists of an optical cavity inside an instantaneous quantum feedback loop. For many of the following calculations, it is advantageous to move into an interaction picture with respect to the free cavity Hamiltonian $H_0 = \hbar \omega_{\rm cav} \, c^\dagger c$. When doing so, we denote state vectors and operators with a subscript ``I." When using the Schr\"odinger picture, state vectors and operators are subscripted with ``S." Here $\omega_{\rm cav} $ is the cavity photon frequency and $c$ denotes the usual cavity photon annihilation operator with $[c,c^\dagger] = 1$.

In the following, we assume that the detection of a photon triggers a very short-time pulse from a laser resonant with the cavity frequency.  If this pulse is always taken from the same laser field, it displaces the field inside the cavity such that the state vector $|\psi_{\rm I}(t) \rangle$ of the cavity in the interaction picture changes into $D(\beta) \, |\psi_{\rm I}(t) \rangle$ where
\begin{eqnarray} \label{Rbeta}
D(\beta) &=& \exp \left( \beta \, c^\dagger - \beta^* \, c \right) 
\end{eqnarray}
and where $\beta$ is a complex number that characterises the strength of the feedback pulse. For simplicity, we assume here that the laser pulse is so strong and short relative to the cavity decay rate $\kappa$ that its effect can be considered instantaneous. In the Schr\"odinger picture, the displacement operator is transformed in the standard way by acting the inverse of the time evolution of the free Hamiltonian $H_0$ upon it.  The result of this is that $\beta$ becomes time dependent and the state vector $|\psi_{\rm S}(t) \rangle$ changes into $D(\beta_{\rm S}(t)) \, |\psi_{\rm S}(t) \rangle$ with 
\begin{eqnarray} \label{Rbeta2}
\beta_{\rm S}(t) &=& {\rm e}^{- {\rm i} \omega_{\rm cav} t} \, \beta \, .
\end{eqnarray}
In other words, we assume here that every feedback pulse is generated by a laser pulse with exactly the same (complex) Rabi frequency.

As mentioned already above, the dynamics of the density matrix $\rho_{\rm I} $ of the cavity in the interaction picture is given by a master equation. In standard form, this equation equals \cite{PRA}
\begin{eqnarray} \label{Cavity master equation feedback0}
\dot{\rho}_{\rm I} & = & - {1 \over 2} \kappa \left[ c^{\dagger} c , \rho_{\rm I} \right]_+ + \eta \kappa \, D(\beta) c \rho_{\rm I} c^{\dagger} D(\beta)^{\dagger} \nonumber \\
&& +  \left(1 - \eta \right) \kappa \, c \rho_{\rm I} c^{\dagger} \, ,
\end{eqnarray}
where $\eta$ is the relevant detector efficiency. Hence the emission of a photon is registered and triggers a quantum feedback pulse with probability $\eta$. Moreover, $1-\eta$ is the probability of an emitted photon leaving the system undetected. The first term in Eq.~(\ref{Cavity master equation feedback0}) models the dynamics of the sub-ensemble without photon emission, while the second and the third term model the dynamics of sub-ensembles with an emission at $t$.

\begin{figure*}[t]
\centering
\includegraphics[width=0.98 \textwidth]{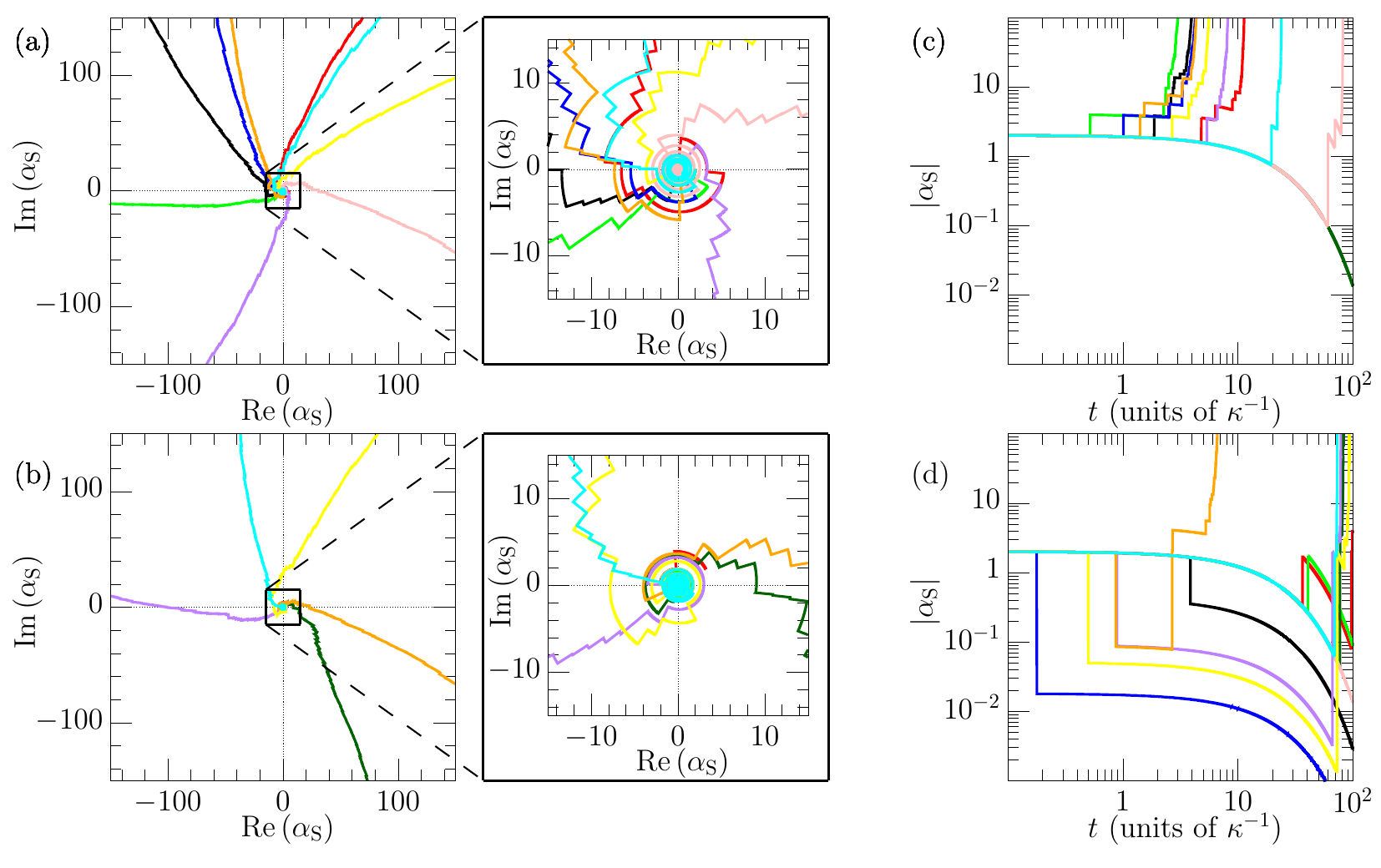}
\caption{[Colour online] Evolution of a random sample of ten quantum trajectories subject to the described unravelling in Eqs.~(\ref{H-cond} - \ref{displace}) according to a quantum jump simulation of total time $t=10 \kappa^{-1}$.  We set $\beta(0) =2$, $\eta=0.5$ and have an initial state with $\alpha(0) = 2$ in (a) and $\alpha(0)=-2$ in (b).  In (a), all but one trajectory diverges from the vacuum state, whereas in (b) only five trajectories diverge.  These tracjectories rapidly separate into the whole phase space, as can be seen more clearly in the Schr\"odinger picture.  The non-diverging trajectories asymptotically approach the vacuum state.  To see this, we plot $|\alpha_{\rm S}(t)|$ in (c) and (d) for (a) and (b) respectively.  Here we see the different times at which various trajectories ``escape'' from the vacuum, while the remaining ones move increasingly closer to this state.}
\label{Individual3-4-together}
\end{figure*}

To predict all possible individual quantum trajectories, we need to unravel Eq.~(\ref{Cavity master equation feedback0}) in a physically meaningful way \cite{Reset,Molmer,Carmichael}. Doing so, we find that under the condition of no photon emission in $(0,t)$, the initial state $|\psi_{\rm I} (0) \rangle$ of the cavity field in the interaction picture evolves with the non-Hermitian conditional Hamiltonian \cite{PRA}
\begin{eqnarray} \label{H-cond}
H_{\rm cond} &=& - {{\rm i} \over 2} \hbar \kappa \, c^{\dagger} c \, .
\end{eqnarray}
This Hamiltonian reduces the length of the initial state vector $|\psi_{\rm I} (0) \rangle$ such that 
\begin{eqnarray} \label{P0}
P_0(t) &=& \left \| \exp \left( - {{\rm i} \over \hbar} H_{\rm cond} t \right) \, |\psi_{\rm I} (0) \rangle \, \right \|^2 
\end{eqnarray}
is the probability for no photon emission in $(0,t)$. Moreover, in the case of an emission at a time $t$, $|\psi_{\rm I} (t) \rangle $ changes (up to normalisation) into $L_1 \, |\psi_{\rm I} (t) \rangle$ or $L_2 \, |\psi_{\rm I} (t) \rangle$, respectively, with $L_1 = c$ and $L_2 = D(\beta) \, c$, depending on whether or not a feedback pulse is triggered. Moreover, the probability density for the emission of a photon at $t$ equals $I(t) = \kappa \, n(t)$ where $n(t) =\langle c^\dagger c \rangle$ is the mean cavity photon number. We now have all the tools needed to generate all the possible quantum trajectories of the resonator in Fig.~\ref{Setup2} numerically.

Before doing so, let us point out that the quantum state of the field inside the optical cavity remains coherent at all times, if the system was initially prepared in a coherent state $|\alpha_{\rm I} (0) \rangle$. For example, under the condition of no photon emission in a time interval $(0,t)$, the cavity field evolves with the conditional Hamiltonian $H_{\rm cond}$ and its (unnormalised state) at time $t$ equals
\begin{eqnarray} \label{xx}
|\psi_{\rm I}^{0} (t) \rangle &=& \exp \left( - {{\rm i} \over \hbar} H_{\rm cond} t \right) \, |\alpha_{\rm I} (0) \rangle \, .
\end{eqnarray}
To simplify this expression, we denote the state with exactly $n$ photons inside the resonator by $|n \rangle$ and take into account that the coherent state $|\alpha_{\rm I} (0) \rangle$ can be written as
\begin{eqnarray} \label{yy}
|\alpha_{\rm I} (0) \rangle &=& {\rm e}^{- {1 \over 2} |\alpha_{\rm I} (0)|^2} \sum_{n=0}^\infty  {\alpha_{\rm I} (0)^n \over \sqrt{n!}} \, |n \rangle \, .
\end{eqnarray}
Applying the time evolution corresponding to the conditional Hamiltonian $H_{\rm cond}$ in Eq.~(\ref{H-cond}) to this state is relatively straightforward. Doing so, we find that 
\begin{eqnarray} \label{yy}
|\alpha_{\rm I} (t) \rangle &=& {\rm e}^{- {1 \over 2} |\alpha_{\rm I} (0)|^2} \sum_{n=0}^\infty  {\left( \alpha_{\rm I} (0) \, {\rm e}^{ - {1 \over 2} \kappa t} \right)^n \over \sqrt{n!}} \, |n \rangle \, . ~~
\end{eqnarray}
After normalisation, $|\psi_{\rm I}^{0} (t) \rangle $ simplifies to the coherent state $|\alpha_{\rm I}(t) \rangle$ with
\begin{eqnarray} \label{exact3}
\alpha_{\rm I}(t) &=& {\rm e}^{- {1 \over 2} \kappa t} \, \alpha_{\rm I} (0) \, .
\end{eqnarray}
Moreover, using Eq.~(\ref{P0}), we find that the probability $P_0(t)$ for no photon emission in $(0,t)$ equals
\begin{eqnarray}
P_0(t) &=& {\rm e}^{ - |\alpha_{\rm I} (0)|^2 \left[ 1 - {\rm e}^{- \kappa t} \right] } \, .
\end{eqnarray}
In the limit of $t \to \infty$, this equation simplifies to the probability $P_0(\infty)$,
\begin{eqnarray} \label{zz}
P_0(\infty) = {\rm e}^{- |\alpha_{\rm S}(0) |^2} \, ,
\end{eqnarray}
which is the probability to never emit a photon given the initial state $|\alpha_{\rm I} (0) \rangle $.

The reason for the damping of $\alpha_{\rm I}(t)$ in Eq.~(\ref{exact3}) is that not emitting a photon reveals information about the cavity state $|\alpha_{\rm I}(t) \rangle$ which needs to be updated accordingly \cite{Reset}. If the emission of a photon goes unnoticed, then the state of the cavity remains unchanged, since the coherent states are the eigenstates of $L_1$. However, if a photon triggers a feedback pulse, the state of the cavity changes from $|\alpha_{\rm I} (t) \rangle$ into 
\begin{eqnarray} \label{displace}
D(\beta) \, |\alpha_{\rm I} (t) \rangle &=& |\alpha_{\rm I} (t) + \beta \rangle
\end{eqnarray}
due to the properties of the displacement operator $D$ in Eq.~(\ref{Rbeta}). Overall, in the interaction picture, each quantum trajectory is fully described by a series of coherent states $|\alpha_{\rm I} (t) \rangle$, which are given by a time dependent complex function $\alpha_{\rm I} (t)$. The same applies in the Schr\"odinger picture, where each quantum trajectory is fully described by the time dependent complex function $\alpha_{\rm S} (t)$ with $\alpha_{\rm S} (t) = \exp (- {\rm i} \omega_{\rm cav} t) \, \alpha_{\rm S} (t) $ and similarly the displacement operator is transformed to $D(\beta_{\rm S}(t))$. Fig.~\ref{Individual3-4-together} shows $\alpha_{\rm S} (t)$ for random samples of individual quantum trajectories which have been generated using standard quantum Monte Carlo simulations \cite{Reset,Molmer,Carmichael}. Figs.~\ref{Individual3-4-together}(a)--(b) show the evolution of ten randomly chosen quantum trajectories, prepared in initial states with $\alpha(0) = 2$ and $\alpha(0) = -2$ respectively.  All other conditions are the same in both setups.  In Figs.~\ref{Individual3-4-together}(c)--(d), we plot for $|\alpha_{\rm s}(t)|$ to show the behaviour of the magnitude of the state more clearly.

\begin{figure}[t]
\centering
\includegraphics[width=0.49 \textwidth]{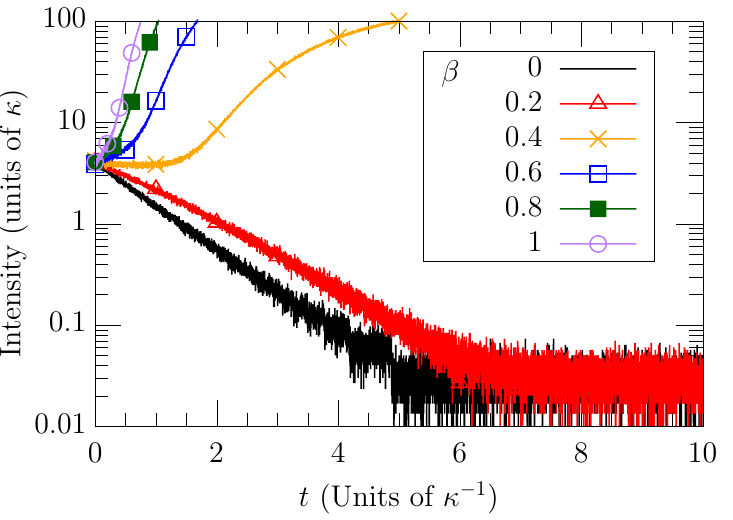}
\caption{[Colour online] Time dependence of the photon emission intensity, defined to be the average number of emissions per unit time, for different feedback parameters $\beta$. As in previous figures, we assume $\alpha_{\rm I}(0) = 2$ and $\eta = 0.5$. The figure is the result of a quantum jump simulation, which averages over $10^6$ individual quantum trajectories. For relatively small values of $\beta$, $I(t)$ tends towards zero. However, as $\beta$ increases, the dynamics of the cavity change and the mean number of photons inside the resonator continues to grow in time.}
\label{Int3}
\end{figure}

\begin{figure}[t]
\centering
\includegraphics[width=0.49 \textwidth]{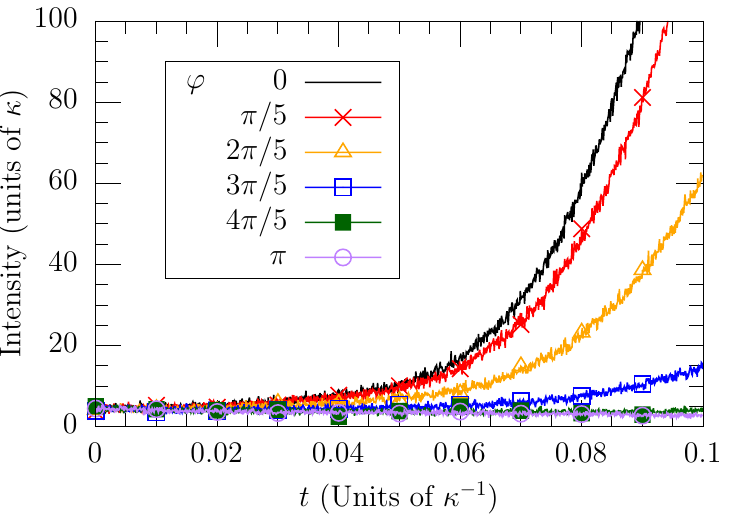}
\caption{[Colour online] Time dependence of the photon emission intensity for different initial states $|\alpha_{\rm I}(0) \rangle$ here with $\alpha_{\rm I}(0) = 2 {\rm e}^{{\rm i} \varphi}$ and varying phase $\varphi$. All other parameters are as in Fig.~\ref{Int3} and results are again generated via a quantum jump simulation averaged over $10^6$ quantum trajectories.  This figure illustrates that there is a strong dependence of the ensemble averages on the initial state of the resonator.}
\label{Int2}
\end{figure}

\section{Non-Ergodicity} A closer look at Fig.~{\ref{Individual3-4-together} shows that there are two very distinct types of dynamics. Some quantum trajectories persistently show an overall increase in the mean number of photons inside the cavity and move further and further away from the vacuum state. In other cases, the mean number of photons inside the resonator remains relatively small, decreases further and the system is extremely likely to eventually reach the vacuum state.  This is not surprising, as photon emissions at a relatively high rate attract more feedback pulses, thereby further increasing the mean number of photons inside the resonator. In the absence of registered emissions and feedback pulses, the no-photon time evolution continuously reduces the field amplitude (c.f.~Eq.~(\ref{exact3})). As we shall see below, the vacuum is the only stationary state of the experimental setup in Fig.~\ref{Setup2}.

Since the cavity field remains always in a coherent state, its density matrix $\rho_{\rm I} (t)$ in the interaction picture always represents a statistical mixture of coherent states. Hence we can assume without restrictions that
\begin{eqnarray} \label{alphass4}
\rho_{\rm I} (t) &=& \int_{C \!\!\!\! I} {\rm d}\alpha \, P(\alpha,t) \, |\alpha \rangle \langle \alpha| \, ,
\end{eqnarray}
where $P(\alpha,t)$ denotes the probability to find a single subsystem of a large ensemble of optical cavities with the same initial state at time $t$ prepared in $|\alpha \rangle$. Substituting this form of the density matrix into the master equation in Eq.~(\ref{Cavity master equation feedback0}) and using the properties of coherent states, we find that 
\begin{eqnarray} \label{alphass4}
\dot{\rho}_{\rm I} (t) &=& \kappa \int_{C \!\!\!\! I} {\rm d}\alpha \, P(\alpha,t) \, \Big[ \eta |\alpha|^2 \, |\alpha + \beta \rangle \langle \alpha + \beta| \nonumber \\
&& + (1 - \eta) |\alpha|^2 \, |\alpha \rangle \langle \alpha| - {1 \over 2} \alpha \, c^\dagger |\alpha \rangle \langle \alpha| \nonumber \\
&& - {1 \over 2} \alpha^* \, |\alpha \rangle \langle \alpha| c \Big] \, . 
\end{eqnarray}
Looking at the various matrix elements of this operator, we find that the cavity possesses a unique stationary state $\rho_{\rm ss}$ with $\dot{\rho}_{\rm I} (t)= 0$. As one would expect, this state is the vacuum state of the resonator, 
\begin{eqnarray} \label{alphass6}
\rho_{\rm ss} &=& |0 \rangle \langle 0| \, .
\end{eqnarray}
When reaching this state the time evolution of the resonator comes to a hold. 

The above-described presence of qualitatively different types of quantum trajectories strongly suggests non-Ergodicity. To show that for sufficiently strong quantum feedback it is not always possible to deduce the statistical properties of the cavity from a single run of the experiment, we now have a closer look at Eq.~(\ref{zz}). This equation shows the probability $P_0(\infty)$ for an optical cavity initially prepared in the coherent state $|\alpha_{\rm I}(0) \rangle$ with $|\alpha_{\rm I}(0)|>0$ to never emit a photon is always positive ($P_0(\infty)>0$) and always smaller than one ($P_0(\infty) < 1$). The larger $|\alpha_{\rm I}(0) |^2$, i.e.~the larger the initial photon number, the more likely it is for the system to experience photon emission. The fact that $P_0(\infty)$ is always less than one shows that there is always a non-zero probability that the cavity field never reaches its vacuum state. The Ergodicity condition of Eqs.~(\ref{ergodicity2}) and (\ref{ergodicity}) being always the same resultantly does not hold for {\em all} possible quantum trajectories.

To show that for sufficiently strong quantum feedback there is indeed always at least one possible quantum trajectory for which the cavity field never reaches the vacuum state, we now have a closer look at the dynamics of the mean cavity photon number $n(t)$. Using the master equation in Eq.~(\ref{Cavity master equation feedback0}), one can show that the time derivative of the expectation value of any observable $A$ averaged over an ensemble of systems equals
\begin{eqnarray} \label{Cavity master equation feedback3}
\langle \dot{A} \rangle & = & - {1 \over 2} \kappa \, \left \langle  \left[ A,c^{\dagger} c \right]_+ \right \rangle
+ \eta \kappa \, \left \langle c^{\dagger}  D(\beta )^{\dagger} A D(\beta) c \right \rangle \nonumber \\
&& + (1-\eta) \kappa \, \left \langle c^{\dagger} A c \right \rangle \, .
\end{eqnarray}
Setting $A = c^\dagger c$, this differential equation can be used to calculate the mean number of photons $n(t)$ inside the cavity. Using Eq.~(\ref{displace}) and the commutator relation $[c,c^\dagger] = 1$ one can show that 
\begin{eqnarray} \label{alphass5}
\dot{n} (t) &=& - \kappa \left[ 1 - \eta \left( |\alpha_{\rm I} (t) + \beta(t)|^2 - |\alpha_{\rm I} (t)|^2 \right) \right] \,  |\alpha_{\rm I} (t)|^2 \, , \nonumber \\
\end{eqnarray}
if the state of the cavity equals $|\alpha_{\rm I} (t) \rangle$. In the absence of any feedback, i.e.~for $\eta = 0$, this equation simplifies to the simple linear differential equation $\dot{n} (t) = - \kappa \, n(t)$, as it should. However, in the presence of sufficiently strong feedback, the $\eta$-term dominates the dynamics of $n(t)$ and the mean cavity photon number evolves in a non-linear fashion, as illustrated in Fig.~\ref{Int3}. 

Suppose the cavity is initially in its vacuum state and a small perturbation moves the resonator into a coherent state $|\alpha_{\rm I} \rangle$ with $|\alpha_{\rm I}|^2 \ll |\beta(t)|^2$. When this applies, Eq.~(\ref{alphass5}) simplifies to a very good approximation to
\begin{eqnarray} \label{alphass7}
\dot{n} (t) &=& - \kappa \, \left( 1 - \eta |\beta(t)|^2 \right) \, |\alpha_{\rm I} |^2 \, .
\end{eqnarray}
with $|\beta(t)|^2 = |\beta|^2$. In the presence of sufficiently strong feedback, i.e.~when 
\begin{eqnarray} \label{alphass77}
\eta \, |\beta|^2 > 1 \, , 
\end{eqnarray}
this time derivative is always positive. When this condition applies, the mean photon number $n(t)$ always grows in time (cf.~Fig.~\ref{Int3}) and the vacuum state becomes a repulsive fixed point of the system dynamics. Qualitatively, the behaviour of the resonator field resembles the behaviour of classical systems with feedback.  This behaviour can be seen in Fig.~\ref{Int3}, where there is a transition from decay to growth of the photon emission intensity as we transition to higher feedback strengths. 

Given its non-Ergodicity and the instability of its stationary state, it is not surprising to also observe a strong dependence of the dynamics of cavity expectation values averaged over an ensemble of optical cavities on the common initial state $|\alpha_{\rm S} (0) \rangle$. In contrast to many other open quantum systems, information about the initial state may persist and may never get lost. This behaviour is illustrated in Fig.~\ref{Int2} which shows the time-dependence of the cavity photon emission rate $I(t)$ for different initial states $|\alpha_{\rm I} (0) \rangle$ with $\alpha_{\rm I} (0) = |\alpha_{\rm I} (0)| \, {\rm e}^{{\rm i} \varphi}$. Although $|\alpha_{\rm I} (0)|$ and the feedback parameter $\beta $ are kept the same, $I(t)$ does not converge to a single value. On the contrary, the distance between curves that correspond to different values of $\varphi$ increases in time. This surprising fact has been shown already to be useful, for example, in quantum metrology protocols where it can be used to overcome the standard quantum limit \cite{PRA,PRA2}.

We further note that the generation of this behaviour is dependent on the existance of what effectively constitutes multiple decay channels, i.e.~those with feedback and those without.  In the case where the channel involving feedback becomes infintesimally weak ($\beta \approx 0$) or is switched off ($\eta \approx 0$), the system returns to the basic Ergodic case of free decay towards the stationary state of the vacuum.  On the other hand, introducing further decay channels by allowing for multiple feedback protocols could lead to even more complex behaviour.

\section{Conclusions} This paper identifies instantaneous quantum feedback as a mechanism to induce non-Ergodicity in open quantum systems. This is achieved by altering the Lindblad operators $L$ of the corresponding master equation (c.f.~Eq.~(\ref{Superoperator2})). As an example, this paper studies an optical cavity inside an instantaneous quantum feedback loop. This system, which remains always in a coherent state, can be analysed relatively easily despite its non-linear dynamics. In the presence of sufficiently strong feedback, the only stationary state of an ensemble of optical cavities with quantum feedback can become repulsive. In general, it is not possible to deduce its dynamics of ensemble averages from individual quantum trajectories.  Moreover, it is shown that the dynamics of ensemble averages can depend strongly on initial states. Since quantum feedback is relatively easy to implement in the laboratory \cite{Feedback}, these properties are expected to find useful applications for the processing of quantum information, for example, when designing quantum versions of Hidden Markov Models \cite{HQMM,HQMM2}. The non-linear dynamics of the experimental setup in Fig.~\ref{Setup2} has already been used to design novel schemes for quantum-enhanced metrology \cite{PRA,PRA2}. Moreover, quantum feedback-induced non-Ergodicity might provide an interesting tool for the design of thermal machines whose efficiency is not restricted by the laws of classical thermodynamics \cite{Jaynes,Jaynes_book}. 

\acknowledgments
LAC acknowledges support from the United Kingdom Engineering and Physical Sciences Research Council (EPSRC) Award No.~1367108 and also Grant No.~EP/P034012/1. Moreover, FT and BM acknowledge support from a University of Leeds UGRL scholarship. In addition, AB was supported by funding from the EPSRC Oxford Quantum Technology Hub NQIT (Grant No.~EP/M013243/1). AB would like to thank Alex Little for stimulating discussions. Statement of compliance with EPSRC policy framework on research data: This publica- tion is theoretical work that does not require supporting research data.


\begin{thebibliography}{100}
\bibitem{Boltzmann}
BIRKHOFF, G. D., {\em Proc. Natl. Acad. Sci. USA} {\bf 17} (1931) 656.

\bibitem{Boltzmann2}
VON NEUMANN, J., {\em Proc. Natl. Acad. Sci. USA} {\bf 18} (1932) 70.

\bibitem{Boltzmann3}
VON NEUMANN, J., {\em Z. Phys.} {\bf 57} (1929) 30.

\bibitem{Kampen}
VAN KAMPEN, N. G., {\em Stochastic processes in Physics and Chemistry} (Elsevier Science B.V., Amsterdam) 1992. 

\bibitem{Mandel}
MANDEL, L. and WOLF, E., {\em Optical coherence and quantum optics} (Cambridge University Press, Cambridge) 1995.

\bibitem{Cresser}
CRESSER, J. D., {\em Ergodicity of Quantum Trajectory Detection Records} in {\em Directions in Quantum Optics} by H. J. Carmichael, R. J. Glauber and M. O. Scully (Editors), Lecture Notes in Physics (Springer-Verlag, Berlin) 2001, p. 358.

\bibitem{Mark}
SREDNICKI, M., {\em Phys. Rev. E} {\bf 50} (1994) 888.
   
\bibitem{many}
POLKOVNIKOV, A., SENGUPTA, K., SILVA, A. and VENGALATTORE, M., {\em Rev. Mod. Phys.} {\bf 83} (2011) 863. 

\bibitem{Mark2}
D'ALESSIO, L., KAFRI, Y., POLKOVNIKOV, A. and RIGOL, M., {\em Adv. Phys.} {\bf 65} (2016) 239.

\bibitem{Broken}
PALMER, R. G., {\em Advances in Physics} {\bf 31} (1982) 669.






\bibitem{Ines}
PARRA-MURILLO, C. A., BRAMBERGER, M., HUBIG, C. and DE VEGA, I., {\em Open quantum systems in thermal non-ergodic environments}, submitted (2020); arXiv:1910.10496.
 
 \bibitem{Papic}  
TURNER, C. J., MICHAILIDIS, A. A., ABANIN, D. A., SERBYN, M. and PAPIC, Z., {\em Nat. Phys.} {\bf 14} (2018) 745.
 
\bibitem{Papic2}
BULL, K., MARTIN, I. and PAPIC, Z., {\em Phys. Rev. Lett.} {\bf 123} (2019) 030601.

\bibitem{Matin}
MATIN, S., PUN, C.-K., GOULD, H., and KLEIN, W., {\em Phys. Rev. E} {\bf 101} (2020) 022103.
        
\bibitem{HQMM}
CLARK, L. A., HUANG, W., BARLOW, T. M. and BEIGE, A., {\em Hidden Quantum Markov Models and Open Quantum Systems with Instantaneous Feedback},
in ISCS 2014: Interdisciplinary Symposium on Complex Systems, Emergence, Complexity and Computation {\bf 14}, (Springer Verlag, Heidelberg) 2015, p. 143; arXiv:1406.5847.

\bibitem{HQMM2}
ADHIKARY, S., SRINIVASAN, S., GORDON, G. and BOOTS, B., {\em Expressiveness and Learning of Hidden Quantum Markov Models}, Proceedings of the 23rdInternational Conference on Artificial Intelligence and Statistics (AISTATS) 2020 (Palermo, Italy) PMLR: Vol. {\bf 108}, 2020; arXiv:1912.02098.

\bibitem{PRA}
CLARK, L. A., STOKES, A. and BEIGE, A., {\em Phys. Rev. A} {\bf 94} (2016) 023840.

\bibitem{PRA2}
CLARK, L. A., STOKES, A. and BEIGE, A., {\em Phys. Rev. A} {\bf 99} (2019) 022102.

\bibitem{Jaynes}
MEHTA, P. and POLKOVNIKOV, A., {\em Ann. Phys.} {\bf 332} (2012) 110.

\bibitem{Jaynes_book}
JAYNES, E. T., {\em Papers on Probability, Statistics and Statistical Physics} (Springer Netherlands) 1989.

\bibitem{Stable1}
CARVALHO, A. R. R. and HOPE, J. J., {\em Phys. Rev. A} {\bf 76} (2007) 010301.

\bibitem{Wiseman-Milburn} 
WISEMAN, H. M. and MILBURN, G. J., {\em Quantum Measurement and Control} (Cambridge University Press) 2010.

\bibitem{Schirmer}
SCHIRMER, S. G. and WANG, X., {\em Phys. Rev. A} {\bf 81} (2010) 062306.

\bibitem{Stable2}
STEVENSON, R. N., HOPE, J. J. and CARVALHO, A. R. R., {\em Phys. Rev. A} {\bf 84} (2011) 022332.

\bibitem{Zelditch}
ZELDITCH, S., {\em in Encyclopedia of Mathematical Physics}, editted by FRAN\c{C}OISE, J.-P., NABER, G. L. and TSUN, T. S. (Academic Press, Oxford) 2006, pp. 183-196.

\bibitem{Lindblad} 
LINDBLAD, G., {\em Comm. Math. Phys.} {\bf 48} (1976) 119.

\bibitem{Reset} 
HEGERFELDT, G. C., {\em Phys. Rev. A} {\bf 47} (1993) 449.

\bibitem{Molmer}
DALIBARD, J., CASTIN, Y. and M{\O}LMER, K., {\em Phys. Rev. Lett.} {\bf 68} (1992) 580.

\bibitem{Carmichael}
CARMICHAEL, H., {\em An Open Systems Approach to Quantum Optics}, Lecture Notes in Physics, Vol.~{\bf 18} (Springer-Verlag Berlin) 1993.

\bibitem{Feedback}
REINDER, J. E., SMITH, W. P., OROZCO, L. A., WISEMAN, H. M. and GAMBETTA, J., {\em Phys. Rev. A} {\bf 70} (2004) 023819.
\end{thebibliography}
\end{document}